\documentclass[prb,twocolumn,showpacs,superscriptaddress,preprintnumbers,amsmath,amssymb]{revtex4}

\usepackage{graphicx}
\usepackage{dcolumn}
\usepackage{bm}
\usepackage{hhline}

\begin{document}

\title{Electronic structures of CeRu$_2X_2$ ($X$ = Si, Ge) in the paramagnetic phase studied by soft X-ray ARPES and hard X-ray photoelectron spectroscopy}

\author{M. Yano}
\author{A. Sekiyama}
\author{H. Fujiwara}
\author{Y. Amano}
\author{S. Imada}
\affiliation{Division of Materials Physics, Graduate School of Engineering  Science, Osaka University, Toyonaka, Osaka 560-8531, Japan}

\author{T. Muro}
\affiliation{Japan Synchrotron Radiation Research Institute, Mikazuki, Sayo, Hyogo 679-5198, Japan}

\author{M. Yabashi}
\affiliation{Japan Synchrotron Radiation Research Institute, Mikazuki, Sayo, Hyogo 679-5198, Japan}
\affiliation{RIKEN, Mikazuki, Sayo, Hyogo 679-5148, Japan}
 
\author{K. Tamasaku}
\author{A. Higashiya}
\author{T. Ishikawa}
\affiliation{RIKEN, Mikazuki, Sayo, Hyogo 679-5148, Japan}

\author{Y. \={O}nuki}
\affiliation{Department of Physics, Graduate School of Science, Osaka University, Toyonaka, Osaka 560-0043, Japan}

\author{S. Suga}
\affiliation{Division of Materials Physics, Graduate School of Engineering  Science, Osaka University, Toyonaka, Osaka 560-8531, Japan}

\date{\today}

\begin{abstract}
Soft and hard X-ray photoelectron spectroscopy (PES) has been performed for one of the heavy fermion system CeRu$_2$Si$_2$ and a $4f$-localized ferromagnet CeRu$_2$Ge$_2$ in the paramagnetic phase.
The three-dimensional band structures and Fermi surface (FS) shapes of CeRu$_2$Si$_2$ have been determined by soft X-ray $h\nu$-dependent angle resolved photoelectron spectroscopy (ARPES).
The differences in the Fermi surface topology and the non-$4f$ electronic structures between CeRu$_2$Si$_2$ and CeRu$_2$Ge$_2$ are qualitatively explained by the band-structure calculation for both $4f$ itinerant and localized models, respectively.
The Ce valences in CeRu$_2X_2$ ($X$ = Si, Ge) at 20 K are quantitatively estimated by the single impurity Anderson model calculation, where the Ce $3d$ hard X-ray core-level PES and Ce $3d$ X-ray absorption spectra have shown stronger hybridization and signature for the partial $4f$ contribution to the conduction electrons in CeRu$_2$Si$_2$.
\end{abstract}

\maketitle
\section{Introduction}
Strongly correlated electron systems show many interesting physical properties due to their complicated electronic structures.
Particularly, Ce based compounds have been extensively studied by both experimental and theoretical approaches because they show variety of $4f$ states due to different hybridization strength between the Ce $4f$ and valence electrons.
Dominance of the RKKY interaction or the Kondo effect is one of the attractive subjects in this system.
Especially, Ce$T_2X_2$ ($T$ = Cu, Ag, Au, Ru, \textit{etc.}; $X$ = Si or Ge) with the tetragonal ThCr$_2$Si$_2$ structure\cite{ThCrSi} (a = b $\sim$ 4 \AA \ and c $\sim$ 10 \AA) shows various $4f$ electron behaviors.\cite{LC,LC2,CMG,Ge_Doniach,HP,RPES}
Among them, the $4f$ electrons in CeRu$_2$Ge$_2$ are localized and ferromagnetically ordered due to ascendancy of the RKKY interaction below the Curie temperature $T_C \sim$ 8 K in ambient pressure.\cite{Ge_TC}
On the other hand, in the case of CeRu$_2$Si$_2$, the Ce $4f$ electron couples with the valence electron and makes the so called Kondo singlet state below the Kondo temperature $T_K \sim$ 20 K.\cite{TK1, TK2} 
CeRu$_2$Si$_2$ is known as a typical heavy fermion system which has a large value of the electronic specific heat coefficient $\gamma \sim$ 350 mJ/mol K$^2$ (about 20 times larger than that of CeRu$_2$Ge$_2$).\cite{Gamma,Gamma1,Gamma2}
Such a difference in $4f$ electronic states is thought to originate from the different hybridization strength between the $4f$ and valence electrons.
For example, when high pressure is applied to CeRu$_2$Ge$_2$, the hybridization strength increases and the electronic structures of CeRu$_2$Ge$_2$ approach to those of CeRu$_2$Si$_2$.\cite{HP,HP2}
Clarification of the electronic structures of CeRu$_2X_2$ is thus the key to reveal a connection between the $4f$ localized and itinerant electronic states.
Therefore, complete study of the electronic states of CeRu$_2X_2$ is essential for understanding physics of the strongly correlated electron systems.

So far, CeRu$_2$Si$_2$ has eagerly been studied in order to elucidate the mechanism of the metamagnetic transition which occurs at $H_m \sim$ 7.7 T.\cite{SiMT,Meta1}
According to the de Haas-van Alphen (dHvA) studies,\cite{dHvA} the Fermi surface (FS) of CeRu$_2$Si$_2$ approaches to that of LaRu$_2$Si$_2$ above $H_m$ due to the localization of the $4f$ electrons.
However, the dHvA results of CeRu$_2$Si$_2$ can not fully be explained by the band structure calculations under $H < H_m$.\cite{dHm2}
For example, the heaviest effective mass FS or the smallest FS signals have not been observed along the magnetic field direction of $<001>$.
Recently, another experimental method named ``soft X-ray $h\nu$-dependent angle-resolved photoelectron spectroscopy (ARPES)'' has been established to determine three-dimensional (3D) FSs by using energy tunable soft X rays from third-generation high brilliance synchrotron radiation light sources.\cite{GeARPES}
In the ARPES studies, we can observe electronic structures in solids at various temperatures and determine the shapes of FSs which can be compared with the results from dHvA measurements.
dHvA measurements can be performed under magnetic fields and high pressures but are confined to low temperatures.
While dHvA measurements probe genuine bulk electronic states, conventional photoelectron spectroscopy (PES) measurements have been believed as a rather surface sensitive technique as far as the photoelectron kinetic energies are in the range of 20 - 200 eV.
However, the bulk sensitivity of the soft X-ray ($h\nu \sim$ 800 eV) PES was confirmed by the observation of the $4f$ PES for Ce compounds.\cite{Nature}
Soft X-ray PES is currently an essential approach to reveal electronic states of transition metal compounds\cite{SRO,Mo} and rare earth compounds.\cite{Yb}

In addition, more bulk-sensitive PES by using hard X rays (HAXPES) has become feasible.\cite{Yamasaki}
The probing depth of PES depends on the kinetic energy of the photoelectron.
According to TPP-2M formula,\cite{TPP2M} a photoelectron inelastic mean free path $\lambda$ can be estimated as a function of the electron kinetic energy ($E_K$).
For example, $\lambda \sim$ 19 \AA \ at $E_K$ = 800 eV and $\lambda \sim$ 115 \AA \ at $E_K$ = 7000 eV for CeRu$_2$Si$_2$.
HAXPES is also useful in order to obtain bulk-sensitive core-level spectra with negligible surface contribution.
For example, the surface spectral weight of the Ce $3d$ level located at the binding energy ($E_B$) of $\sim$ 900 eV can be significantly reduced by HAXPES.

By virtue of the soft X-ray ARPES experiments, we have so far clarified 3D FSs of CeRu$_2$Ge$_2$ in the paramagnetic phase.
The results of the ARPES measurements were compared with the LDA calculation for LaRu$_2$Ge$_2$ performed on the Ce $4f$ electron localized model.\cite{GeLDA}
The difference between the ARPES results and the calculation for LaRu$_2$Ge$_2$ or dHvA results for CeRu$_2$Ge$_2$ in the ferromagnetic phase\cite{King} can be explained by non-negligible small hybridization between its Ce $4f$ and valence electrons.\cite{GeARPES}
We have extended the study to $h\nu$-dependent soft X-ray ARPES, HAXPES and X-ray absorption spectroscopy (XAS) for CeRu$_2$Ge$_2$ and a heavy fermion system CeRu$_2$Si$_2$ in order to reveal their electronic structures.
The ARPES results for CeRu$_2$Si$_2$ are compared with those for CeRu$_2$Ge$_2$ in the paramagnetic phase and the band-structure calculation for CeRu$_2$Si$_2$, in which the $4f$ electrons are treated as itinerant.
The HAXPES and XAS spectra have been analyzed by the single impurity Anderson model (SIAM), by which the clear differences in the mean $4f$ electron number and hybridization strength between CeRu$_2$Si$_2$ and CeRu$_2$Ge$_2$ are confirmed.
We show the transformation of 3D FSs resulting from different hybridization strength between Ce $4f$ and valence elections.

\section{Methods}
\subsection{Soft X-ray ARPES and Ce $3d$ XAS}
The CeRu$_2X_2$ single crystals were grown by the Czochralski pulling method.\cite{CP}
The soft X-ray ARPES and XAS measurements were performed at BL25SU\cite{SR25} in SPring-8.
A SCIENTA SES200 analyzer was used covering more than a whole Brillouin zone along the  direction of the slit.\cite{angle}
The energy resolution was set to $\sim $ 200 meV for FS mappings and $\sim$ 100 meV for a high resolution measurement.
The angular resolution was $\pm$0.1\r{} and $\pm$0.15\r{} for the perpendicular and parallel directions to the analyzer slit, respectively.
These values correspond to the momentum resolution of $\pm$0.025 \AA$^{-1}$ and $\pm$0.038 \AA$^{-1}$ at $h\nu$ = 800 eV.
The clean surface was obtained by cleaving \textit{in situ} providing a (001) plane in the base pressure of $\sim 3\times 10^{-8}$ Pa.
All of the ARPES measurements were performed at 20 K.
The surface cleanliness was confirmed by the absence of the O $1s$ photoelectron signals.
We have measured Pd valence band to determine the Fermi level ($E_F$) and estimate the energy resolution of the system.
The measurements of the valence band and the Si $2p$ core spectra were alternated for the purpose of normalization of each valence band spectrum.
In ARPES measurements, we have first performed the $k_z-k_{xy}$ mapping at several $h\nu$ and angles.
Photon momenta were taken into account to determine the exact value of $|k_z|$.\cite{photon}
In order to analyze ARPES data as functions of the binding energy and momentum, we have employed both energy distribution curves (EDCs) and momentum distribution curves (MDCs).
The XAS was measured by the total electron yield mode whose probing depth is comparable to that of the HAXPES.
The energy resolution was set to better than 200 meV.
The detailed experimental conditions are given in Ref. \onlinecite{XAS}.

\subsection{Hard X-ray photoelectron spectroscopy (HAXPES)}
\begin{figure}[htbp]
  \begin{center}
    \includegraphics[keepaspectratio=true,height=80mm]{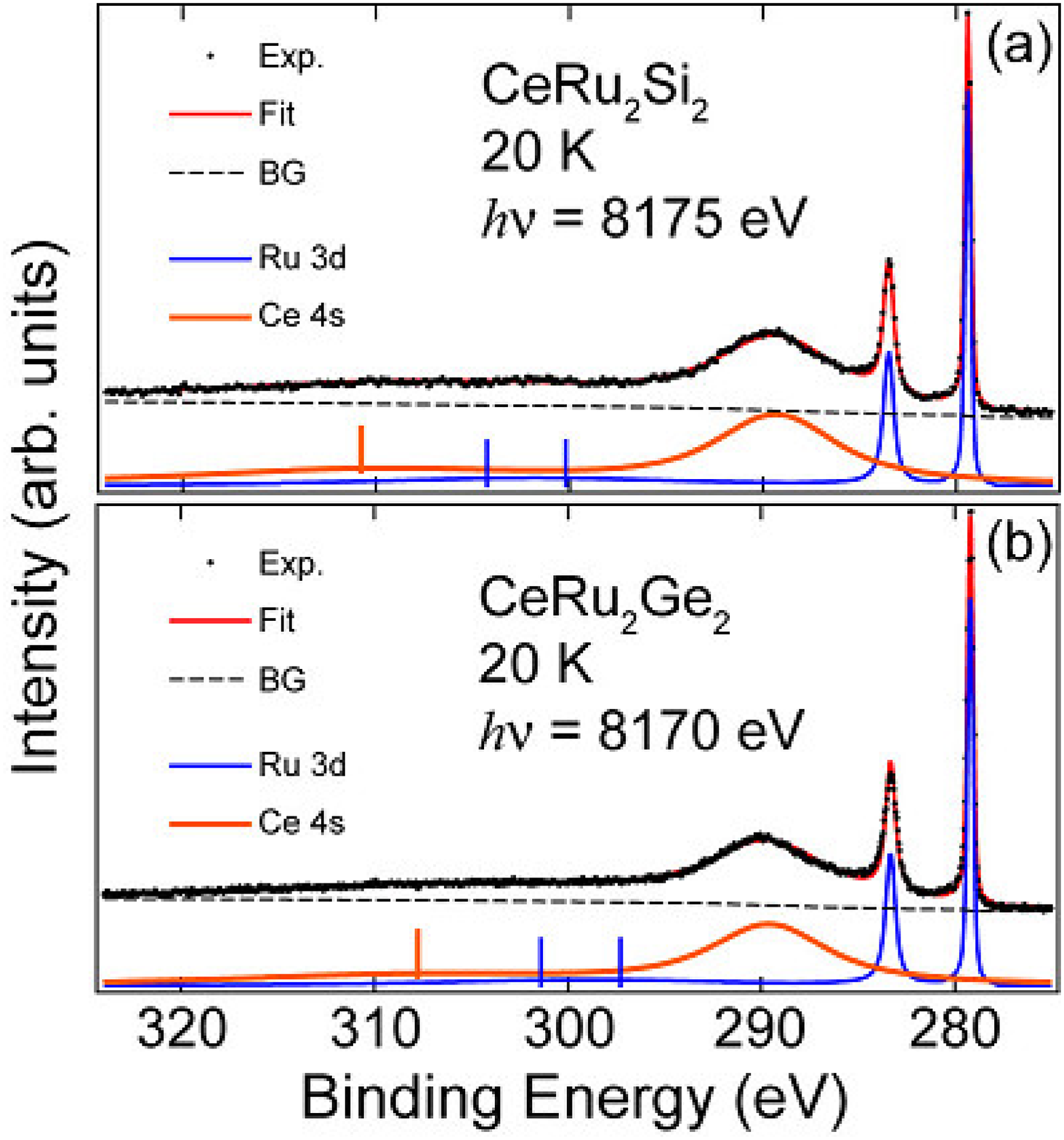}
  \end{center}
  \caption{ (color online). Ce $4s$ and Ru $3d$ core-level HAXPES spectra for (a) CeRu$_2$Si$_2$ and (b) CeRu$_2$Ge$_2$. The spectra were fitted by Gaussian and Lorentian broadenings of the Mahan's line shape\cite{Mahan} for each component and plasmon. The dashed line represents background (BG) of each spectrum. The energy loss peak appears at about 20 eV higher binding energy from each main peak as shown by vertical bars.}
  \label{Plasmon}
\end{figure}

The HAXPES measurements for CeRu$_2X_2$ were carried out at BL19LXU\cite{SR19} in SPring-8 with a MB Scientific A1-HE analyzer.
The (001) clean surface was obtained by cleaving \textit{in situ} in the pressure of $10^{-8}$ Pa at the measuring temperature of 20 K.
The photon energy was set to about 8 keV and the energy resolution was set to about 400 meV for the Ce $3d$ core-level measurements.
We have measured an evaporated Au valence band spectrum to determine $E_F$, and the Ce $4s$ and Ru $3d$ spectra to estimate the energy loss (including plasmon excitation) peak position and its intensity.
Figure \ref{Plasmon} shows the HAXPES spectra of the Ce $4s$ and Ru $3d$ core-levels for CeRu$_2$Si$_2$ and CeRu$_2$Ge$_2$ with the energy resolution of about 200 meV.
In CeRu$_2$Si$_2$ spectra, the sharp Ru $3d_{5/2}$ and $3d_{3/2}$ levels locate at 279.4 eV and 283.5 eV, respectively.
The rather broad Ce $4s$ state lies at 289.5 eV.\cite{Cross}
Rather weak and broad energy loss peaks are located about 20.8 eV from the main peak for CeRu$_2$Si$_2$ and about 18.1 eV for CeRu$_2$Ge$_2$ according to a line shape analysis.\cite{Mahan}
The energy loss peak position from the main peak and its intensity were later taken into account in the fitting of the Ce $3d$ PES spectra.

\subsection{Single impurity Anderson model (SIAM)}
In order to obtain the information on the bulk $4f$ electronic states as well as the mean $4f$ electron number ($n_f$) from the Ce $3d$ core-level PES and XAS for CeRu$_2X_2$, we performed the SIAM calculation based on the $1/N_f$-expansion method developed by Gunnarsson and Sch{\"o}nhammer.\cite{GS}
Here, $N_f$ (the degeneracy of the Ce $4f$ level) was set to 14 for simplicity.
We calculated the $3d$ PES and XAS spectra to the lowest order in $1/N_f$, where the $f^0$, $f^1$, and $f^2$ configurations were taken into account for the initial state.
We divided the band continuum into $N$ (we set to 21) discrete levels following Jo and Kotani.\cite{JoKotani}
The configuration dependence of the hybridization strength was also taken into account, and was chosen to be the same as that obtained for $\alpha$-Ce by Gunnarsson and Jepsen.\cite{Edep,iiwake1}
The energy dependence of the hybridization strength was assumed to be constant in the binding energy range from 0 ($E_F$) to $B$ (we set to 4 eV).
The multiplet effects were not taken into account for simplicity.\cite{iiwake2}

\section{Results}
\subsection{3D ARPES}
\begin{figure}[htbp]
  \begin{center}
    \includegraphics[keepaspectratio=true,height=55mm]{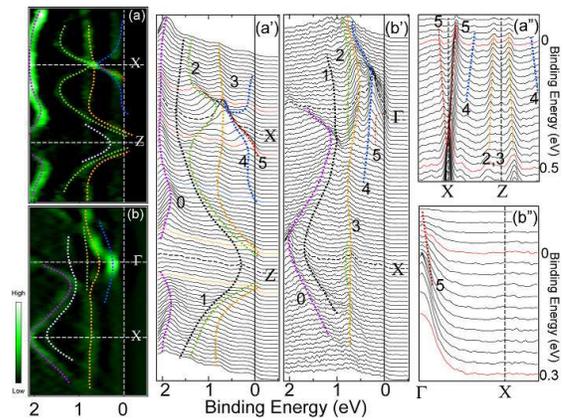}
  \end{center}
  \caption{ (color online). ARPES spectra of CeRu$_2$Si$_2$ near $E_F$ at 20 K with $h\nu$ = 725 eV. (a) and (b) are the second order differential images along the Z-X and $\Gamma$-X directions. EDC (a') and (b') cover the same regions of (a) and (b). (a") is MDCs of the same region as (a) or (a'). (b") is the expanded MDCs along the $\Gamma$-X direction. The energy resolutions of (a) and (b) series are set to about 200 meV and 100 meV, respectively. The dashed lines representing each band are guides to the eye.}
  \label{EDC}
\end{figure}

Figures \ref{EDC} (a), (b) and (a'), (b') display the soft X-ray ARPES results for CeRu$_2$Si$_2$, indicating existence of six bands in the region from $E_F$ to about 2 eV.
These bands are numbered from 0 to 5 from the higher binding energy side.
According to Figs. \ref{EDC} (a) and (a') along the Z-X direction, the energy positions of the bands 2 to 5 approach the X point, and therefore a strong intensity peak is observed at about 0.69 eV.
The band 1 has the lowest-binging energy at the Z point and does not cross $E_F$ while the bands 2 and 3 cross $E_F$ near the Z point as a merged band due to the limited resolution.
The bands 4 and 5 have been observed separately as clearly shown in the expanded MDCs (a") and these bands can be then traced in Fig. \ref{EDC} (a').
Figures \ref{EDC} (b) and (b') along the $\Gamma$-X direction show some bands, whose dispersions are smaller than those along the Z-X direction except for the band 0.
Along the $\Gamma$-X direction, the bands 0 to 4 are on the occupied side while only the band 5 has Fermi wave number ($k_F$) near the $\Gamma$ point as shown in the expanded figure (b"), although the strong intensity of the band 4 is overlapped near the $\Gamma$ point.
Since the contribution from the band 4 is relatively decreased when the energy approaches $E_F$, the contribution of the band 5 is confirmed.
The band 5 can be then traced in Figs. \ref{EDC} (b').

\begin{figure}[htbp]
  \begin{center}
    \includegraphics[keepaspectratio=true,height=50mm]{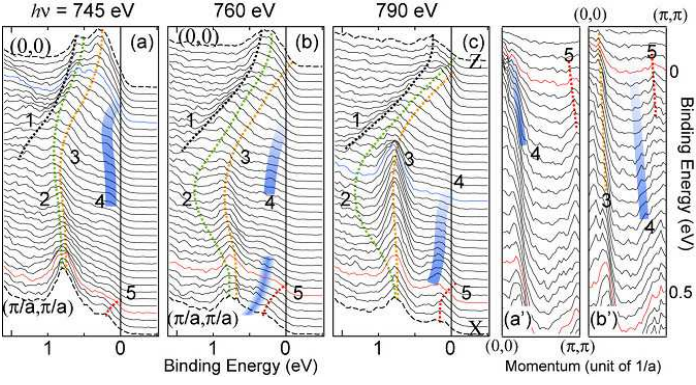}
  \end{center}
  \caption{ (color online). $h\nu$-dependent ARPES spectra of CeRu$_2$Si$_2$ along the (0,0) - ($\pi /a$, $\pi /a$) direction at 20 K. The values of $k_z$ are (a) $\frac{4}{13} \frac{2\pi}{c}$ and (b) $\frac{7}{13} \frac{2\pi}{c}$ achieved by the photon energies of 745 eV and 760 eV, respectively. (c) corresponds to the high symmetry line Z-X achieved by 790 eV ($k_z=\frac{2\pi}{c}$). (a') and (b') are the expanded MDC figures corresponding to (a) and (b), respectively. The dashed lines representing each band are guides to the eye. The thick lines are given for the band 4 whose intensities are weaker than others.}
  \label{kz_EX}
\end{figure}

Figure \ref{kz_EX} shows the $h\nu$-dependent ARPES results along the (0, 0) - ($\pi /a$, $\pi /a$) direction.
The shape of each band and $k_F$ have comprehensively been evaluated by both EDCs and MDCs.
As shown in Figs. \ref{kz_EX} (a) and (a'), the band 5 crosses $E_F$ near the ($\pi /a$, $\pi /a$) point.
Although the spectral weight derived from the band 4 is weak, we can trace the band 4 dispersion crossing $E_F$.
Figures from (a) to (c) also indicate that the $k_F$ for the band 4 approaches to ($\pi /a$, $\pi /a$) point when $h\nu$ is away from 745 eV toward 790 eV although the intensity of the band 4 at $h\nu$ = 760 eV is so weak that $k_F$ cannot be determined so accurately.
Therefore the position of the band 4 is given by the zones to allow possible experimental ambiguity.
Hereafter the lines of the guides to the eye are drawn to pass through the adjacent (in the sense of wavenumber in EDCs and energy in MDCs) noticeable structures in order to compromise with the experimental statistics.
When the excitation photon energy is increased from $k_z \sim 0$ ($h\nu$ = 725 eV) as (a) $\to $ (b) $\to$ (c), we can recognize that the bands 2 and 3 cross $E_F$ sequentially.
At $h\nu$ = 745 eV ($k_z = \frac{4}{13}\frac{2\pi}{c}$), the bands 1, 2 and 3 are on the occupied side at the (0, 0) point.
When $k_z = \frac{7}{13}\frac{2\pi}{c}$ was chosen by 760 eV (b), only the band 3 crosses $E_F$ near the (0,0) point while the bands 1 and 2 are fully occupied.
At $h\nu$ = 790 eV ($k_z = \frac{2\pi}{c}$), both bands 2 and 3 cross $E_F$ near the Z (0, 0, $2\pi /c$) point while the band 1 is still on the occupied side at the binding energy of 0.27 eV, indicating that the band 1 does not form FS.

\begin{figure}[htbp]
  \begin{center}
    \includegraphics[keepaspectratio=true,height=70mm]{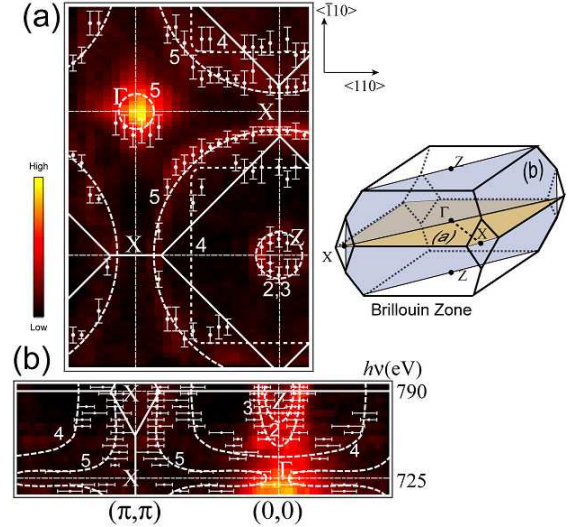}
  \end{center}
  \caption{ (color online).  FSs slices of CeRu$_2$Si$_2$ at 20 K obtained by integrating the photoelectron intensity from 0 eV to -0.1 eV, and the Brillouin zone for the ThCr$_2$Si$_2$-type structure. The solid lines represent the corresponding Brillouin zone and the dashed-and-dotted lines represent high symmetry lines. White dots with error bars represent the estimated $k_F$. The dashed lines represent the FSs following the experimentally evaluated $k_F$s. (a) FSs slice in the $k_x-k_y$ plane at $k_z \sim 0$ ($h\nu $= 725 eV). (b) FSs slice in the $k_z$(ordinate) - $k_{xy}$(abscissa) plane. $h\nu =$725 eV and 790 eV correspond to the $\Gamma $ point and the Z point, respectively, along the $\Gamma $-Z direction.}
  \label{FSall}
\end{figure}

We have determined $k_F$s by means of both EDCs and MDCs for the FS mapping.
We have integrated the intensities of MDCs from $E_F$ to -0.1 eV as a function of momentum from a slice of the ARPES data.
The topology of the FSs thus obtained is displayed in Fig. \ref{FSall}.
Figure \ref{FSall} (a) shows a $k_x-k_y$ slice at $k_z \sim 0$ obtained by changing the detector angles and (b) shows a $k_{xy}-k_z$ slice including the $\Gamma$-X axis whose $k_z$ corresponds to the excitation photon energies from 715 eV to 805 eV with 5 eV steps.
The $k_F$s estimated from EDCs and MDCs are plotted by dots on the figure with error bars.
As shown in Fig. \ref{FSall} (a), there is a small hole-like FS centered in the vicinity of the Z point derived from the bands 2 and 3.
The contour of the hole-like FS derived from the band 4 exists mostly inside the square Brillouin zone centered at the Z point.
Its intensity in the vicinity of $E_F$ is considerably small compared with that for CeRu$_2$Ge$_2$,\cite{GeARPES} suggesting that the electron correlation in the band 4 is larger for CeRu$_2$Si$_2$ than for CeRu$_2$Ge$_2$ because the intensity of the coherent part of a band is suppressed by the smaller magnitude of the quasiparticle renormalization factor (or coherent factor) due to electron correlations.\cite{Damascelli,renormal}
The largest circle shaped electron FS of the band 5, whose obvious contour is clearly seen, centered at the Z point surrounds the square Brillouin zone.
Some intensities derived from the band 5 centered at the $\Gamma$ point can also be seen.
The strongest intensity in Fig. \ref{FSall} (a) around the $\Gamma$ point is caused by the spectral weight of the band 4 near $E_F$ which does not, however, cross $E_F$ near the $\Gamma$ point as shown in Fig. \ref{EDC} (b').
In Fig. \ref{FSall} (b) are shown some outlines of FSs in the $k_z-k_{xy}$ plane.
The elliptical contours centered at the Z point  derived from the bands 2, 3 can be separately observed as in Fig. \ref{kz_EX}.
The prolonged elliptical contour of the FS derived from the band 4 can also be confirmed in Fig. \ref{FSall} (b) along the in-plane Z-X direction.
Another contour of FS derived from the band 5, which is symmetric with respect to the $k_z$ axis of X-X or Z-$\Gamma$ and has a narrow part near the $\Gamma$ point, is also seen in Fig. \ref{FSall} (b).

\begin{figure}[htbp]
  \begin{center}
    \includegraphics[keepaspectratio=true,height=85mm]{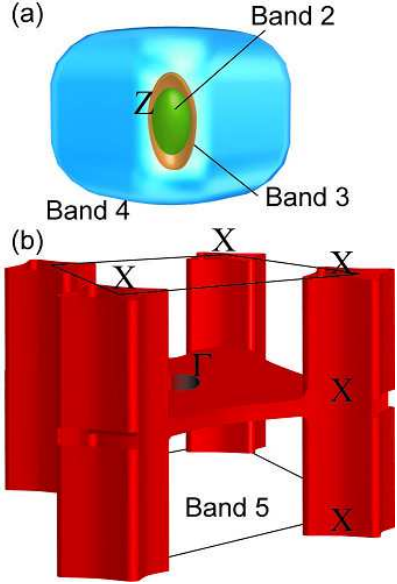}
  \end{center}
  \caption{(color online). Qualitative 3D FSs images of CeRu$_2$Si$_2$ at 20 K obtained by $h\nu$-dependent soft X-ray ARPES. The shapes of the 3D FSs were determined by the results in Figs. \ref{FSall} (a) and (b). The obtained hole-like FSs derived from the bands 2, 3 and 4 centered at the Z point are shown in (a). FSs from the bands 2 and 3 are in the FS from the band 4 and FSs of the bands 3 and 4 are intentionally opened up here so that the inner FS can be visible. The band 5 constructs a rather complicated FS shape as shown in (b). The center of (b) corresponds to the $\Gamma$ point.}
  \label{3D}
\end{figure}

\begin{figure}[htbp]
  \begin{center}
    \includegraphics[keepaspectratio=true,height=80mm]{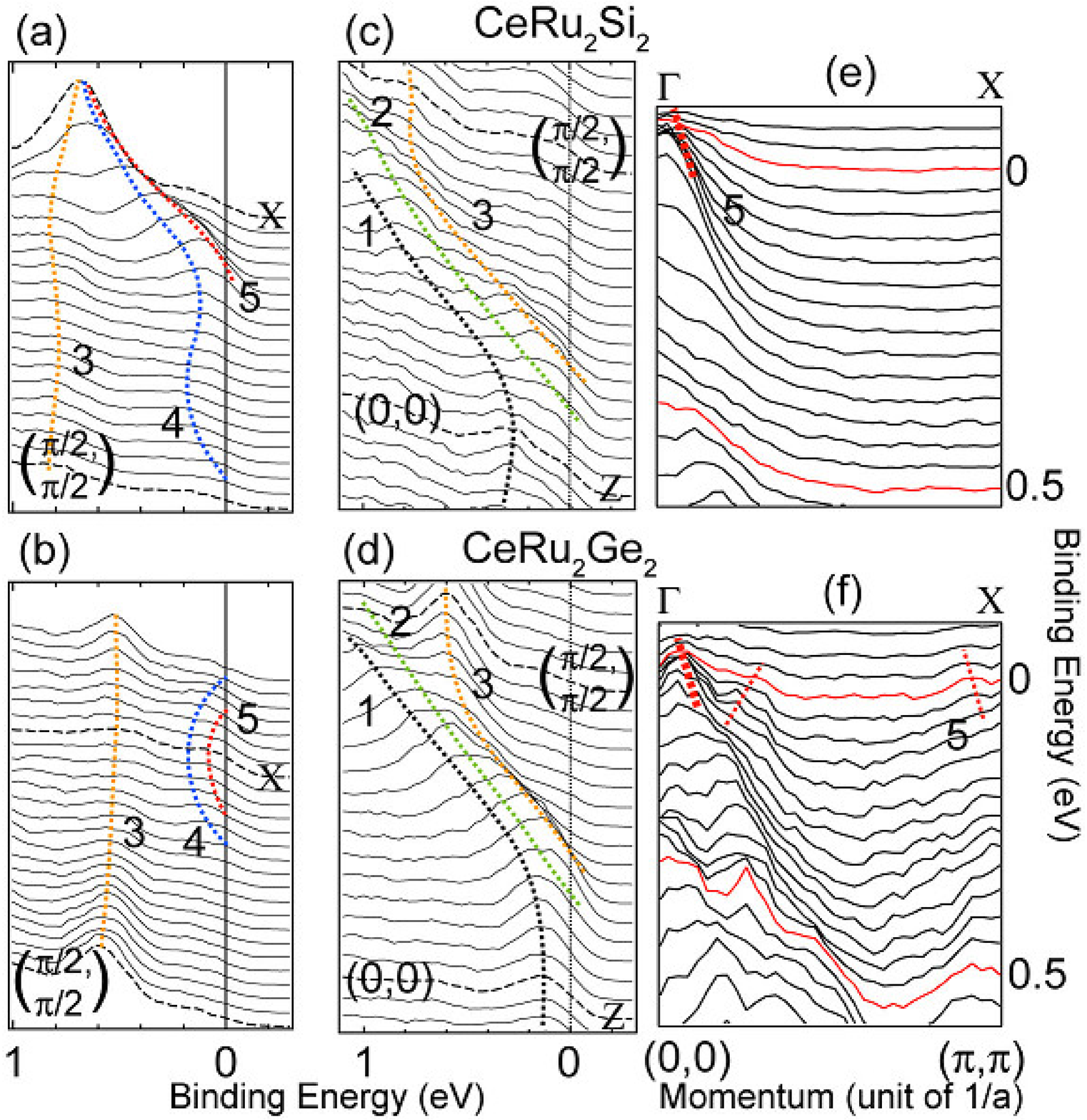}
  \end{center}
  \caption{(color online). The expanded EDC- and MDC- displayed ARPES spectra near $E_F$ of CeRu$_2X_2$ at 20 K. Upper figures are for CeRu$_2$Si$_2$ and lower figures are for CeRu$_2$Ge$_2$ in the paramagnetic phase, respectively. The numbered dashed lines which represent respective bands are guides to the eye. (a)-(d) are EDCs along the Z-X direction at $k_z \sim 2\pi/c$. The photon energies are (a), (c) 725 eV and (b), (d) 755 eV, respectively. (e) and (f) are MDCs along the $\Gamma$-X direction at $k_z \sim 0$. The photon energies are (e) 725 eV and (f) 820 eV, respectively. The details for CeRu$_2$Ge$_2$ can be seen in Ref. \onlinecite{GeARPES}}
  \label{Detail}
\end{figure}

From these two slices of the FSs, we suggest rough 3D shapes of the FSs of CeRu$_2$Si$_2$ in Fig. \ref{3D}.
It was found that CeRu$_2$Si$_2$ has four FSs derived from the bands 2 to 5.
The bands 2 and 3 form ellipsoidal shaped hole-like FSs prolonged along the $k_z$ direction centered at the Z point.
The prolonged length of the FS of the band 2 is shorter than that of the band 3 as confirmed by Fig. \ref{kz_EX} and Fig. \ref{FSall} (b).
The FS of the band 2 is surrounded by that of the band 3.
The band 4 forms a large hole-like swelled-disk FS centered at the Z point.
This FS encompasses both FSs of the bands 2 and 3.
These hole like FSs are similar to those of the ARPES results for CeRu$_2$Ge$_2$ in the paramagnetic phase, although the size of the FS of CeRu$_2$Si$_2$ derived from the band 4 is smaller than that of CeRu$_2$Ge$_2$ (Figs. \ref{Detail} (a), (b)).\cite{GeARPES}
Meanwhile, the shape of the FS formed by the band 5 is quantitatively different between CeRu$_2$Si$_2$ and CeRu$_2$Ge$_2$.
Namely the FS from the band 5 for CeRu$_2$Si$_2$ can be understood as if the small doughnut-like FS surrounding the $\Gamma$ point for CeRu$_2$Ge$_2$ in the paramagnetic phase  expands and touches the cylindrically shaped FS along the $k_z$ direction centered at the X point.
The detailed difference of the band structures between CeRu$_2$Si$_2$ and CeRu$_2$Ge$_2$ are shown in Fig. \ref{Detail} and will be discussed below.

\subsection{$3d$ core-level HAXPES and XAS}

\begin{figure}[htbp]
  \begin{center}
    \includegraphics[keepaspectratio=true,height=50mm]{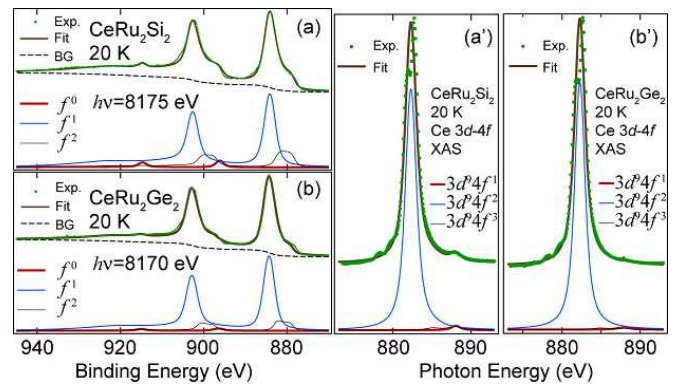}
  \end{center}
  \caption{(color online). Ce $3d$ HAXPES spectra ($h\nu $ = 8175 eV and 8170 eV) and the fitted results; (a) CeRu$_2$Si$_2$ and (b) CeRu$_2$Ge$_2$. The Ce $3d-4f$ XAS spectra\cite{XAS} and fitted results; (a') CeRu$_2$Si$_2$ and (b') CeRu$_2$Ge$_2$. The dots represent the experimental data and the solid lines in the upper part of each figure are fitted results. The dashed lines are the background (BG). The deconvoluted $f^n$($f^{n+1}$) final states in the $3d$ HAXPES (XAS) spectra components are also shown in the bottom of each figure.}
  \label{SiGe_PES}
\end{figure}

In order to clarify the bulk Ce $4f$ states in CeRu$_2X_2$, we have performed the Ce $3d$ core level HAXPES and Ce $3d-4f$ XAS.
Figures \ref{SiGe_PES} (a) and (b) show HAXPES results at 20 K for CeRu$_2$Si$_2$ and CeRu$_2$Ge$_2$, respectively.
The $f^0$ and $f^2$ photoelectron emission final state components in CeRu$_2$Si$_2$ spectrum are stronger than in the spectrum of CeRu$_2$Ge$_2$, whose $f^0$ contribution is very small.
In Figs. \ref{SiGe_PES} (a') and (b'), XAS results for CeRu$_2X_2$ are shown.
A shoulder structure around 887.8 eV can be seen in CeRu$_2$Si$_2$ spectrum while it is very weak for CeRu$_2$Ge$_2$.
This shoulder structure is mainly represented by the $f^1$ XAS final state component.
These differences in the spectra between CeRu$_2$Si$_2$ and CeRu$_2$Ge$_2$ reflect the relatively itinerant $4f$ character for CeRu$_2$Si$_2$.

In order to estimate the $4f$ electron number $n_f$ and $f^n (n = 0,1,2)$ contributions in the initial state for CeRu$_2$Si$_2$ and CeRu$_2$Ge$_2$, we have fitted both $3d$ core-level HAXPES and XAS spectra by the SIAM calculation with unique parameter sets.
The optimized parameters in the calculation are the bare $4f$ binding energy ${\epsilon}_f$, the $4f-4f$ on-site Coulomb repulsive energy $U_{ff}$, the $4f-$core-level Coulomb attractive energy $U_{fc}$, and the hybridization strength $V$ defined by $\sqrt{N}v$ ($N$ is 21 in this calculation), where $v$ is the hybridization strength between the $4f$ and one discrete level (same definition as in Ref. \onlinecite{JoKotani}).
The mean hybridization strength often used for the SIAM calculations \cite{AllenOh1986} defined by $\Delta \equiv ({\pi}/B)\int_0^B{\rho}v^2(E)dE$, where ${\rho}v^2(E)$ is the energy dependence of the hybridization strength between the $4f$ level and continuum valence band, can be evaluated as $\sim ({\pi}/B)V^2$ ($B$ is 4 eV in this calculation).

The results of the SIAM calculation are summarized in Fig. \ref{SiGe_PES} and Table \ref{table}.
As shown in Fig. \ref{SiGe_PES}, the SIAM calculation well reproduces the experimental spectra by using the unique parameter set for each compound.\cite{multiplet}
The optimized parameters are comparable to those in the NCA calculation for the bulk $4f$ photoemission spectra.\cite{RPES}
The estimated $n_f$ is very close to 1 for CeRu$_2$Ge$_2$, reflecting its localized $4f$ character.
Still there is a tiny amount of ``non-$f^1$'' contributions in the initial state for CeRu$_2$Ge$_2$ in the paramagnetic phase at 20 K.
The shift of $E_F$ compared with that for LaRu$_2$Ge$_2$ seen in the ARPES results for CeRu$_2$Ge$_2$ is thought to be attributable to these non-$f^1$ components.
$n_f$ is also found to be close to 1 for CeRu$_2$Si$_2$.
However, the $f^0$ contribution is about twice larger and the $f^2$ weight is apparently larger than for CeRu$_2$Ge$_2$.
The $f^1$ contribution in the initial state is less than 0.9 but considerably larger than those for such strongly valence-fluctuating systems as Ce$T_2$ ($T =$ Fe, Rh, Ni, and Ir)\cite{KonishiCeFe2,AllenOh1986,Jung2003}.
We conclude that this ``in-between'' value of the initial $f^1$ weight reflects the heavy fermion behavior and metamagnetic transition (localization of the $4f$ state) under high magnetic fields for CeRu$_2$Si$_2$.

\begin{table*}[htbp]
 \caption{Optimized parameters ($\epsilon_{4f}$, $U_{ff}$, $U_{fc}$, and $V$ given in units of eV) and estimated $f^0,f^1$ and $f^2$ contributions and $n_f$ in the initial state for CeRu$_2X_2$ by the SIAM calculation.}
 \begin{ruledtabular}
  \begin{tabular}{ccccccccc}
       & $\epsilon_{4f}$  & $U_{ff}$  & $U_{fc}$ & $V$ & $f^0$ &  $f^1$ & $f^2$ & $n_f$  \\
    \hline
     CeRu$_2$Si$_2$  & 1.7  & 7.0  & 10.6  & 0.295  &  0.060 & 0.894 & 0.047 & 0.987  \\
   
     CeRu$_2$Ge$_2$ & 1.7  & 7.0  & 10.6  & 0.234  & 0.028 & 0.942 & 0.030 & 1.002   \\
  \end{tabular}
  \end{ruledtabular}
 \label{table}
\end{table*}

\section{Discussion}
The dHvA measurements for CeRu$_2$Si$_2$ have shown four FSs,\cite{dHm2} from which the effective mass of each FS has been estimated.
The effective mass has been reported as $1.5m_0$ for the FS of the band 2, $1.6m_0$ for the FS of the band 3, $120m_0$ for the FS of the band 4 and $10-20m_0$ for the FS of the band 5 ($m_0$ : mass of a free electron).\cite{dHm1}
If a band had the heavy electron character, the slope of the band dispersion close to $E_F$ would be very small and spectral weight near $E_F$ would be weak.
Thus, the slope of the band 4 is thought to be small from the dHvA results since the band 4 has the heaviest effective mass.
The smaller slope of the band 4 near $E_F$ has been observed in our ARPES measurements as revealed in Figs. \ref{EDC} and \ref{kz_EX}.
Furthermore, intensity of the band 4 is weaker than any other observed bands due to smaller renormalization factor.
The ARPES results suggest that the effective mass of the band 4 is largest among the whole bands forming FSs, being consistent with the dHvA results.

\begin{figure}[htbp]
  \begin{center}
    \includegraphics[keepaspectratio=true,height=40mm]{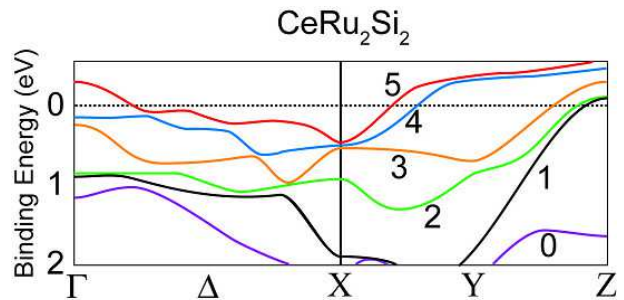}
  \end{center}
  \caption{(color online). The band calculation with APW method for CeRu$_2$Si$_2$\cite{SidHvA} along the $\Gamma$-X and Z-X direction within a region between -0.5 eV and 2 eV. Only the bands from 0 to 5 are displayed.}
    \label{LDA02}
\end{figure}

Apart from the band 4, most of the band structures of CeRu$_2$Si$_2$ revealed by the soft X-ray ARPES results resemble those obtained by the LDA calculation for CeRu$_2$Si$_2$\cite{SidHvA} which treats $4f$ electrons as itinerant.
The band calculation for CeRu$_2$Si$_2$\cite{SidHvA} is shown in Fig. \ref{LDA02}.
When we compare Fig. \ref{EDC} of the ARPES results and Fig. \ref{LDA02} of the LDA calculation, the qualitative consistency between the experimental results and the band calculation is recognized.
The shape of each band along the Z-X direction obtained by ARPES is similar to that by the calculation.
The approach of the four bands 2 to 5 toward the X point is remarkably similar to the calculation which predicts the approach of the bands 3 to 5 at the X point.
However, it was experimentally found that the band 1 does not cross $E_F$ as shown in Fig. \ref{Detail} (c) in spite that the band calculation\cite{SidHvA,Runge} predicts that the band 1 crosses $E_F$ near the Z point.
Furthermore, the absence of the FS derived from the band 1 in our ARPES is consistent with the results of dHvA measurement for CeRu$_2$Si$_2$.

In Figs. \ref{Detail} (c) and (d) are shown the expanded figures of EDCs near the Z point for CeRu$_2X_2$.
The peak position of the band 1 at the Z point for CeRu$_2$Si$_2$ is located at higher-binding energies than that of CeRu$_2$Ge$_2$.
The band 1's peak position of CeRu$_2$Si$_2$ at the Z point is 0.27 eV, whereas that of CeRu$_2$Ge$_2$ is about 0.14 eV.
Additionally, Figs. \ref{Detail} (a) and (b) show the difference between the bottom positions of the band 4 and the band 3 in CeRu$_2X_2$.
The position of each band of CeRu$_2$Si$_2$ lies in higher binding energies than those of CeRu$_2$Ge$_2$ near the X point.
The prominent difference of the band structures between CeRu$_2X_2$ appears for the band 5 as shown in Figs. \ref{Detail} (e) and (f).
This difference of the band 5 obtained from ARPES is in good agreement with the difference of the band calculations between LaRu$_2$Ge$_2$ and CeRu$_2$Si$_2$.\cite{SidHvA,GeLDA}
The band 5 of LaRu$_2$Ge$_2$ crosses $E_F$ three times in the region along the $\Gamma$-X direction, while that of CeRu$_2$Si$_2$ crosses only once in the same region.
When we take a different perspective, it is possible to think that $E_F$ of CeRu$_2$Si$_2$ is shifted to lower binding energies or to the unoccupied side compared to CeRu$_2$Ge$_2$ along the $\Gamma$-X direction.

These differences of the band position in CeRu$_2X_2$ is roughly understood if $E_F$ of CeRu$_2$Si$_2$ is energetically higher than that of CeRu$_2$Ge$_2$ in the paramagnetic phase.
The $E_F$ shift of CeRu$_2$Si$_2$ from CeRu$_2$Ge$_2$ in the paramagnetic phase is caused by the increased number of electrons contributing to the bands forming the FSs in CeRu$_2$Si$_2$ due to the hybridization with the $4f$ electrons.
As confirmed by the SIAM calculation of the $3d$ PES and XAS spectra, and by the $3d-4f$ resonance PES, the hybridization is stronger for CeRu$_2$Si$_2$ than for CeRu$_2$Ge$_2$ indicating the $E_F$ shift mentioned above.

For more precise understanding of the difference in the electronic structures between CeRu$_2$Si$_2$ and CeRu$_2$Ge$_2$, this ``rigid-band-like" energy shift is not sufficient because the band structures themselves can be modified due to the different hybridization strength at different $k$ values.
Indeed, such a modification can be seen in Figs. \ref{Detail} (a), (b) and (c), (d).
For instance, the bands 3 to 5 are nearly degenerated at the X point in CeRu$_2$Si$_2$, whereas these are energetically separated in CeRu$_2$Ge$_2$.
The band 5-dispersion along the $\Gamma$-X direction shown in Figs. \ref{Detail} (e) and (f), especially near the $\Gamma$ point, is also essentially modified because the simple $E_F$ shift would lead to the shift of $k_F$ due to the band 5 indicated by the bold dotted line in Figs. \ref{Detail}(e) and \ref{Detail}(f) toward the $\Gamma$ point for CeRu$_2$Si$_2$, which is inconsistent with the experimental results.
In this way, the essential differences in the band structures between CeRu$_2$Si$_2$ and CeRu$_2$Ge$_2$ are experimentally clarified.

\section{Conclusion}
We have performed bulk-sensitive 3D ARPES Ce $3d$ core-level HAXPES and XAS for a heavy fermion system CeRu$_2$Si$_2$ and a $4f$-localized system CeRu$_2$Ge$_2$ by using soft and hard X rays.
The detailed band structures and the shapes of the FSs of CeRu$_2$Si$_2$ are revealed and they are found to be different from those of CeRu$_2$Ge$_2$.
The differences between them are consistent with the differences between the calculation for $4f$ electron itinerant model and localized one.
The FS shapes of CeRu$_2X_2$ are consistently understood as the reflection of the hybridization strength between the Ce $4f$ and valence electrons, which is revealed by the analysis based on SIAM for the Ce $3d$ HAXPES and $3d-4f$ XAS spectra.
Each mean $4f$ electron number of CeRu$_2X_2$ is quantitatively estimated in agreement with the qualitative changes in FSs.

\section*{Acknowledgment}
We are grateful to H. Yamagami for fruitful discussions.
We thank J. Yamaguchi, T. Saita, T. Miyamachi, H. Higashimichi, and Y. Saitoh for supporting the experiments.
The soft X-ray ARPES was performed under the approval of the Japan Synchrotron Radiation Research Institute (Proposal Nos. 2004A6009, 2006A1167, 2007A1005).
This work was supported by the Grant-in-Aids for Scientific Research (15G213, 1814007, 18684015) of MEXT, Japan, and the 21st Century COE program (G18) of JSPS, Japan.
This work was also supported by the Asahi Glass Foundation and Hyogo Science and Technology Association.


\begin{thebibliography}{51}
\bibitem{angle}The detectable angle along the slit is about 10\r{}. $|\Gamma-X|$ distance of CeRu$_2$Si$_2$ is $\sqrt{2}\pi/a \sim$1.06 \AA$^{-1}$ in a reciprocal space. The angle corresponding to this distance is about 4.2\r{} at $h\nu =$ 800 eV. Thus the analyzer can cover whole Brillouin zone.
\bibitem{photon}If an X ray was incident onto a sample at 45\r{} with respect to the surface normal, for example, this incident photon has the momentum parallel $(q_{||})$ and perpendicular $(q_{\bot})$ to the surface. When $h\nu \sim$ 800 eV, the photon momentum values of both $|q_{||}|$ and $|q_{\bot}|$ are about 0.29 \AA$^{-1}$.
\bibitem{Cross}The Ce $4s$ intensity is relatively stronger than that of the Ru $3d$ excitations because the relative (Ce $4s$)/(Ru $3d$) cross section is large at $h\nu$ = 8180 eV ($\sim $1.5) compared with that at such conventional soft X-ray excitations as $h\nu$ = 1486 eV ($\sim $0.1).\cite{Yeh}
\bibitem{iiwake1}The configuration dependence of the hybridization strength is essentially important in order to simultaneously reproduce both core-level PES and XAS spectra by using the same parameters\cite{KonishiCeFe2}.
\bibitem{iiwake2}In order to well reproduce the spectra by the SIAM calculation without considering the multiplet effects, we have employed the Gaussian broadening with different energy widths for the $f^0$ ($f^1$), $f^1$ ($f^2$), and $f^2$ ($f^3$) configurations. The width for the $f^1$ contribution in the core-level photoemission spectra was chosen to be larger than that for the other contributions.
\end{thebibliography}
\end{document}